# Stochastic Earned Value Analysis using Monte Carlo Simulation and Statistical Learning Techniques

**Fernando Acebes** [a]

**María Pereda** [b]

**David Poza** [a]

**Javier Pajares** [a]

**José Manuel Galán** [b]

[a]INSISOC, Departamento de Organización de Empresas y Comercialización e Investigación de Mercados, Escuela de Ingenierías Industriales, Universidad de Valladolid, Paseo del Cauce 59, Valladolid 47011, Spain

[b]INSISOC, Área de Organización de Empresas, Departamento de Ingeniería Civil, Escuela Politécnica Superior, Universidad de Burgos, Edificio La Milanera, C/Villadiego S/N, Burgos 09001, Spain

**Abstract**

*The aim of this paper is to describe a new integrated methodology for project control under uncertainty. This proposal is based on Earned Value Methodology and risk analysis and presents several refinements to previous methodologies. More specifically, the approach uses extensive Monte Carlo simulation to obtain information about the expected behavior of the project. This dataset is exploited in several ways using different statistical learning methodologies in a structured fashion. Initially, simulations are used to detect if project deviations are a consequence of the expected variability using Anomaly Detection algorithms. If the project follows this expected variability, probabilities of success in cost and time and expected cost and total duration of the project can be estimated using classification and regression approaches.*

Simulation, Project Risk Management, Statistical Learning, Anomaly detection.

## 1. Introduction

Project control and monitoring involve comparing a plan or baseline with the actual performance of the project. The analyses of these deviations are aimed at taking actions, if needed, to early correct the possible problems that can put in danger the objectives of the plan. The most popular managerial methodology used in Project Management is Earned Value Management (EVM). This framework integrates in a unified approach, three dimensions of the project –scope, time and cost– using monetary units as common pivotal measure (Abba and Niel, 2010, Anbari, 2003, Blanco, 2013, Burke, 2003, Cioffi, 2006, Fleming and Koppelman, 2005, Henderson, 2003, Henderson, 2004, Jacob, 2003, Jacob and Kane, 2004, Kim et al., 2003, Lipke, 2003, Lipke, 2004, Lipke, 1999, McKim et al., 2000).

Recent research enhances the standard approach to EVM incorporating statistical analysis, learning curves or fuzzy set theory, especially for project predictions at completion (Colin and Vanhoucke, 2014, Hazir, 2014, Lipke et al., 2009, Moslemi Naeni and Salehipour, 2011, Naeni et al., 2011, Narbaev and De Marco, 2014, Plaza and Turetken, 2009, Tseng, 2011, Wauters and Vanhoucke, 2014). An active area of development is currently focused on integration of EVM with risk management analysis. Progress in this line has produced decision tools based on two metrics to estimate if the deviations may be caused by structural problems or if they are compatible with the expected range of variability, derived from the stochastic nature of the project –estimated variability of costs and durations of project activities- (Pajares and López-Paredes, 2011). These results have been refined using Monte Carlo simulation and statistical control charts (Acebes et al., 2014).

In this work, we initially investigate the alternative use of Anomaly Detection algorithms to detect structural deviations in projects. Assuming the stochastic definition of the project, we then advance the research proposing the use of statistical learning techniques and Monte Carlo simulation to estimate the probability of over-runs (delays or over-costs) and the success decision boundaries. The analysis is completed with additional methodologies to predict not only over-runs but also the expected budget and time.

The rest of the paper is organized as follows. First, in the 2. Background section, we review previous methodologies related to the same problem we face in this research. Then in the 3. Methodology section, we explain the statistical learning methodologies we use in our framework, and how we can apply them for project control. In particular, we explain the role of Anomaly Detection Algorithms, and Classification and Regression techniques. Finally, we address a case study to show how these methodologies work together with EVM for project control.

## 2. Background

EVM does not provide a way to determine whether deviations are due to the expected range of variability associated with the stochastic nature of the project or they may be caused by unexpected events affecting project runtime. Knowing the reasons for project over-runs would provide the project manager with valuable information for decision-making. Concretely, the fact of the deviations exceeding the expected variability would be a warning sign that the project realization is probably not running as planned. Being aware of this situation would allow applying actions to redirect the evolution of the project.

With this idea in mind, previous research has provided two different frameworks based on EVM that inform whether deviations are within the probabilistic expected level or not: the Schedule Control Index (SCoI) / Cost Control Index (CCoI) Framework (Pajares and López-Paredes, 2011) and, more recently, the Triad Methodology (Acebes et al., 2014).

### 2.1 The Schedule Control Index (SCoI) and Cost Control Index (CCoI) Framework

Pajares and Lopez-Paredes (2011) use Monte Carlo simulation to obtain the statistical distribution of the cost and the duration at the end of the project. This information is used to select the confidence level (both in terms of time and in terms of cost) that will be used to monitor the performance of the project). If the cost (time) at the end of a particular project is below the cost (time) at the selected confidence level, the cost overrun (delay) is considered to be caused by the randomness of the real costs (durations) of the activities. Therefore, the difference between the project cost (duration) at the confidence level and the mean project cost (duration) gives an idea of the maximum deviation that can be explained by the stochastic nature of the cost (duration) of the activities. In other words, this difference is considered as the size of the cost / time buffers for the project.

Nevertheless, knowing the size of these buffers at the very end of the project is useless since it does not permit to make decisions that rectify the project performance during runtime. It would be desirable, however, to know what portion of these buffers is available at any particular time during the project execution. To this aim, the authors develop a system to redistribute the size of project cost and time buffers throughout the project life cycle. In order to determine the portion of buffer that corresponds to each period, the authors define the Project Risk Baseline as the residual uncertainty to complete the remaining activities of the project. Then, every time interval is provided with a portion of the cost / time buffers namely $ACBf_t$ and $ASBf_t$ respectively. The size of these buffers are proportional to the risk eliminated between two consecutives periods (i.e. the difference between two adjacent points in the risk baseline).

The authors define two control indices based on these buffers: Cost Control Index (CCoI) and Schedule Control Index (SCoI), which are equal to the traditional indices used by EVM: Schedule Variance (SV) and Cost Variance (CV) plus the corresponding portion of the cost / time buffer $ACBf_{t=ES}$ / $ASBf_t$:

$$SCoI_t = ASBf_t + SV$$

$$CCoI_t = ACBf_{t=ES} + CV$$

Therefore, the new criteria to diagnose the time performance of the project are the following: the project is behind schedule if SV < 0 (as in the traditional EVM). However, depending on the value of $ASBf_t$, this delay may be due to by the randomness of the real duration of the activities or caused by structural problems. If $ASBf_t$ is greater than SV (and thus $SCoI_t > 0$), we infer that the delay falls within the expected variability. However, if SV is greater than $ASBf_t$ (and thus $SCoI_t < 0$), the delay may be caused by structural problems and thus require measures to redirect the performance of the project.

Similarly, the index $CCoI_t$ warns about cost overruns (CV > 0) and, when they occur, it reports whether the overruns are within the expected variability ($CCoI_t > 0$) or not ($CCoI_t < 0$).

### 2.2 The Triad Methodology: (x, t, c)

Acebes et al. (2014) developed a different method to determine whether the project deviations are within the expected variability or whether, on the contrary, they are due to undiscovered factors affecting the project performance. This method also uses Monte Carlo simulation to obtain the statistic distribution of all the possible realizations of a project. However, unlike the method shown above, the authors directly determine the statistical distribution of cost and time at intermediate percentages of completion of the project.

For every realization of a Monte Carlo simulation, the system provides a final cost and time. That is, when the percentage of completion of a simulation is 100% (x = 100%), we obtain the final cost for that simulation ($c_{100\%}$) and the final duration for that simulation ($t_{100\%}$). This triad (100%, $t_{100\%}$, $c_{100\%}$,) gives a name to this methodology. Afterwards, the algorithm calculates the cost and time at the desired intermediate time intervals for that particular simulation. It is important to mention that the percentage of completion of the project is calculated in terms of cost. This means that, for example, the project is 50% completed at the time when the cost of that realization of the simulation reached a half of its final value. Therefore, at this point of the simulation, the triad that defines the state of the project is (50%, $t_{50\%}$, $c_{50\%}$).

Each realization of the Monte Carlo simulation will reach the intermediate points of completion at different levels of cost and time. This leads to a point cloud (in terms of time and cost) at the selected intermediate points. This information allows calculating statistical information at these intermediate time intervals. Therefore, we can determine a confidence level that may be used during the realization of the real project to check if the actual values of cost and time at that percentage of completion fall within the confidence level (project under control) or not (need for corrective measures).

In order to determine if the cost / time of a running project is under control (or not) at any time for a given confidence interval, the authors represent the values of the triads in two dimensional graphs: one for cost monitoring (x, $c_x$) and another for time monitoring ($t_x$, x).

## 3. Methodology

### 3.1 Triads and Monte Carlo simulation

The basis of this research stems from the triad method by Acebes et al. (2014) described in the previous section. In the same way, we also generate a data universe (realizations of the project) by means of Monte Carlo simulation, and this data is used to find the statistical properties of the project at any point during its execution.

However, we make a change in the variables making up the triad. Specifically, we consider a triad comprised of the terms (EV, t, c) instead of (%, t, c). That is to say, for every realization of the Monte Carlo simulation, we register the values of EV (earned value) along with the corresponding cost (c) at several intermediate points (t) along the project life cycle. This refinement overcomes some of the limitations of the technique proposed by Acebes et al. (2014), as it required assuming that EV was linear with work execution (i.e. EV = % · BAC; BAC: Budget At Completion) whereas the methodology proposed in this article does not require such a hypothesis. Furthermore, the new approximation is more usable and intuitive from a practitioners' viewpoint. In fact, when monitoring a project it is not straightforward to calculate the accurate percentage of completion; whereas EV (i.e. the budgeted cost of work performed) is a concept project managers are familiar to work with.

Therefore, by means of Monte Carlo simulation, we obtain - at certain levels of EV along the project execution - a point cloud that represents the time and the actual cost of every realization of the simulation. Then, advanced statistical methodologies are applied to this data. On the one hand, treating the data as a classification problem allows knowing whether the project will finish in time and cost. On the other hand, processing the data as a regression problem allows forecasting the expected cost and time at termination of the project.

### 3.2 As a 2D density distribution. Anomaly Detection algorithm

Acebes et al. (2014) try to discern the ranging of expected variability generating different instances of the project according to the planned variability in different development stages. These simulation results are used to build percentile curves to estimate the expected ranges of costs and time. This approach represents an improvement compared to previous techniques but ignores the time-cost correlation that is sometimes implicit in the definition of duration and cost of individual activities. Decoupling these magnitudes can prevent from detecting anomalous situations in the project (producing false negatives). One of the contributions of this work is to propose the techniques developed for anomaly detection (also known as novelty detection, outlier detection, deviation detection or exception mining (Ding et al., 2014)) together with Monte Carlo simulation and triad methodology as support tool to control and monitor stochastic projects (estimating if project deviations are consequences of the project variability) and - if needed - to prompt the appropriate correcting actions.

Novelty detection entails finding the observations in test data that differ in some respect from the training data (Pimentel et al., 2014). Typically the problem emerges in situations in which there are enough data from normal events but data about abnormal situations is rare or inexistent. Precisely the goal of a solution to this problem is identifying observations that deviate or are inconsistent with the sample data in which they occur (Hodge and Austin, 2004).

In order to identify situations during development of the project abnormal to the stochastic definition of the project, again multiple instances of the project are simulated using the methodology described in the previous section. Given a development stage of the project determined by an earned value, the idea is to build a model that describes the normal range of behavior of the project. This normality model is used as a test, comparing it with the development of the real project. A wide range of the techniques to solve this problem gives then a score (sometimes probabilistic) to the given observation that compared with a decision threshold results in a report about the abnormality of the situation. Basically, if the deviations of the project can be explained by its planned variability or if they are indicating abnormal performance and the need to take correction measures.

There are several families of strategies to deal novelty detection techniques (Hodge and Austin, 2004, Markou and Singh, 2003a, Markou and Singh, 2003b, Pimentel et al., 2014). Given their interpretability, in this work we use a probabilistic approach. These methodologies are focused on estimating the generative probability density function from the training data. This function is then used to calculate the probability that a new observation may have been generated by the distribution (Pimentel et al., 2014). Taking into account that the amount of data is obtained by simulation (and consequently can be large enough to have low variance) we have used a non-parametric approach – multivariate density estimation– to reduce bias.

Multivariate density estimation involves fitting a surface, a *kernel*, on every point of the data set and smoothing its contribution into the space around. Then all surfaces are aggregated together giving the overall density function. This process is summarized in Fig. 1

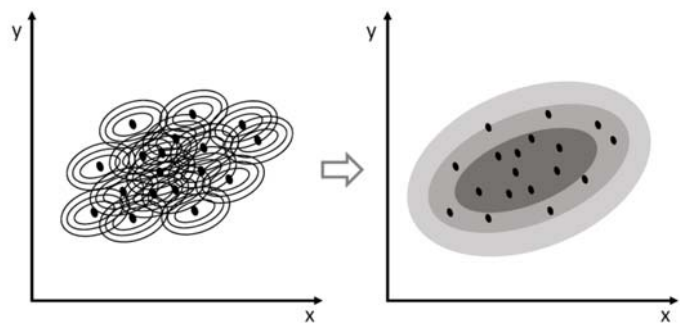

*Fig. 1. 2D kernel density estimation approach. Every point in the dataset includes an individual kernel (left), all the individual kernels are aggregated to obtain a general density function (right). Based on http://en.wikipedia.org/wiki/Multivariate_kernel_density_estimation*

The selection of the smoothing parameter, bandwidth, is crucial because it determines the quantity of smoothing of the kernels and so the accuracy of the fitting. In multivariate density estimation, the bandwidth is a matrix that allows the orientation of the kernels and so the orientation of the estimated density surface. Different techniques for bandwidth selection have been widely studied for the univariate case, but the best methodology is yet considered case-dependent (Zambom and Dias, 2012). For the multivariate problem, the most advanced selector to date is the smooth cross-validation (SCV) methodology, which fits a full bandwidth matrix (previous techniques adjusted diagonal bandwidth matrices) (Duong and Hazelton, 2005).

We have used the function *Hscv* for the smoothed cross-validation (SCV) bandwidth selector and the function *kde* for kernel density estimation with radial kernels, both from the R package *ks* (Duong, 2007).

### 3.3 As a classification problem

Analyzing the data as a classification problem allows us to estimate the probability of the project of finishing in time and/or in cost. The following sections are devoted to explaining what we understand by classification and the type of validation used to select a classification technique and estimate its error. The different classifier algorithms used are then succinctly presented..

#### *3.3.1 Classification*

A classification problem aims to predict a quantitative variable, often called *response, output* or *dependent variable*, with a set of qualitative and/or quantitative variables called *predictors*, *independent variables*, or just *variables*. Adopting this perspective –a classification problem– with the simulated triads can be very useful for project control analysis. Let's assume a stochastic monitored project in any intermediate situation, a relevant question for a project manager could be: "will the project finish in time and cost?" Or more precisely, "how likely is the project to finish in time and cost?". If completing the project in time is modelled as a quantitative variable Yes/No or completing the project under budget is also modelled as quantitative variable Yes/No, all the ingredients for a classification problem are present.

Using the EVM framework, any project in progress can be characterized by its t=AT (actual time), its AC (actual cost) and its percentage of completion $x_{AT}=EV_{AT}$ / BAC. We propose using classification algorithms (classifiers) that fed with this information can provide an answer to the previous questions. In order to do so, our methodology involves selecting an appropriate classifier and, for each x of analysis (when x is equal to $x_{AT}$), training the classifier with the Monte Carlo simulations. For each simulation j and an x of analysis, $T_{xj}$ and $C_{xj}$ are the independent variables of the classification problem to train the classifier, and the output of each instance is the state of the project from the budget perspective, i.e., if the j simulation finished under budget or not; or the state of the project from the time perspective, i.e., if it finished in time or not. Once the classifier is trained, the output of the monitored project can be predicted given its current AT, AC and EV. Beyond that, taking into account that many classification algorithms do not only compute the most likely output but also the probability of each response, the classifier can also provide the probability of finishing in time and under budget if no additional correction measures are taken. Moreover, project managers can also be reported with the decision boundaries of the algorithm, that is, the lines AT, AC for a given intermediate x that partitions the space of development of the project in more chances to finish in time or cost than to incur in delay or over-cost. Or in other words, what are the stochastic limits of time, cost in the development of the project that maintain a favorable forecast for the project.

#### *3.3.2 Nested Cross Validation*

Classification algorithms often capture patterns that are characteristic of the particular training data set but that cannot be generalized to independent data, especially in high flexible and low biased models. Classifiers *overfit* training data, and hence we can only trust in those assessments based on data not used during training.

Cross-validation techniques group together a family of strategies to properly select the specific model among several options or assess the model's performance of a given choice. K-fold cross-validation is one of the most popular approaches. This process consists in randomly partitioning a data set into k subsets –folds– with the same number of elements. Correlatively the classifier is trained using k-1 folds and assessed with the remaining fold as independent data. Averaged results of k rounds decrease the influence of any particular training/test data division. Moreover, multiple measures of the performance allow estimating the dispersion and the test error of the statistic. Previous research shows that k=5 or k=10 provide an adequate bias-variance trade-off of models assuming a moderate computational cost (Hastie et al., 2009).

Although cross-validation can be used for model selection and model assessment, recent studies (Anderssen et al., 2006, Varma and Simon, 2006) warn about an optimistic bias in error estimation when the error obtained during model selection is reported as an estimation of the model performance. In order to obtain an unbiased estimation of the true performance error a nested cross-validation scheme has been suggested (Anderssen et al., 2006, Varma and Simon, 2006).

Stone (1974) summarizes the gist of nested cross-validation as "cross-validatory assessment of the cross-validatory choice". Nested cross-validation involves two nested loops: an inner loop aimed at model selection in which the parameters of the algorithm are estimated, and an outer loop to assess the unbiased performance of the model selected in the inner loop. Original data is partitioned using a k-fold cross-validation scheme. The inner loop receives iteratively data from k-1 folds which in turn are employed as input data in a k-fold analysis for every combination of parameters tested for model selection. The model with lower error in the inner fold is tested in the outer loop.

### 3.3.3 Quadratic Discrimination Analysis

One of the most popular Bayesian classifiers is the discriminant analysis. Given an observation *x*, these algorithms are based on computing the probabilities *Pr(Y = k|X = x)* for each possible class *k*, and assigning the observation to the class with higher probability. In order to estimate these conditional probabilities (posterior probabilities) Bayes' theorem is used:

$$\Pr(Y = k|X = x) = \frac{\pi_k f_k(x)}{\sum_{l=1}^{k} \pi_l f_l(x)} \qquad (1)$$

Where $\pi_k$ is known as *prior* probability and denotes the probability that a randomly chosen observation belongs to the class *k* and $f_k(x)$ is the density function of *X* for an observation that belongs to the class *k* ($Pr(X = x|Y = k)$). When the training data contains enough observations and comes from a random sample, $\pi_k$ can be estimated as the proportion of each class. However, obtaining an estimation of $f_k(x)$ requires some assumptions about the form of the distribution. Quadratic discriminant analysis (QDA) assumes that each $f_k(x)$ follows a multivariate Gaussian distribution with a class-specific mean and a class-specific covariance matrix (Hastie et al., 2009). This assumption entails estimating more parameters than if we suppose that the covariance matrix is equal for all the classes (Linear discriminant analysis), however this a better choice since QDA is more flexible –boundaries separating classes can be any conic section and consequently nonlinear– and the size of the training set is not a major concern for this application. Besides, this classifier directly gives a probability of belonging to each k class, something relevant from a risk management point of view. A project manager controlling a project in an intermediate stage is not only interested if finishing a project on time is more probable than completing it with delay but also is relevant to estimate the probabilities of each event. We have used R package "MASS" (Venables and Ripley, 2002) for the QDA classification analysis in this work.

### 3.3.4 Random Forest

A tree is a nonparametric model composed of nodes and links in a hierarchical structure that can be used for classification or regression. In a classification problem, every node represents a test and terminal nodes associate an input that passes all tests until that terminal node, with a class label.

Tree models present many advantages. They are invariant to monotonic transformations, robust and straightforward to interpret. But they can suffer from high variance problems, that is, the tendency to overfit.

Random Forests are a technique for reducing variance in high-variance low bias machine learning methods (Breiman, 2001)). Based on the concept of bagging and bootstrap aggregation (Breiman, 1996)), random forests consist on building an ensemble of models, trees, to form a super model, the *forest*. Each tree is built from an uncorrelated bootstrap sample from the training data set, using in each node only a subset of the predictors available to decide. Once all the trees are trained, they are all grouped in a combined metric, such as a majority rule vote in the case of a classification problem.

Random forest are becoming widely used by reason of its advantageous features (Criminisi et al., 2011): they have fast and parallelizable algorithms, do not suffer from overfitting, and can exploit "Out-Of the Bag" (OOB) data to analyze the relative importance of the predictors in the classification decision and to estimate the classification error pretty accurately.

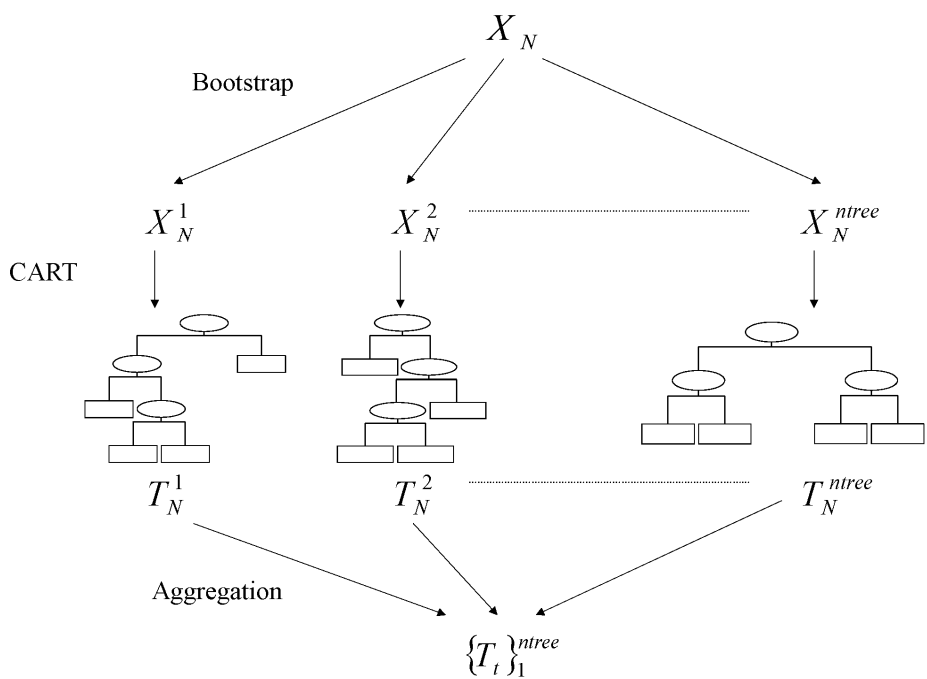

*Fig. 2. The random forest algorithm generates ntree CART trees (T), each one trained from a bootstrapped sample X and using stochastically selected features at each split. For classification problems each tree of the forest (ensemble) votes to give a consensual class.*

### 3.3.5 Support Vector Machines

Support Vector Machines (SVM) are supervised learning algorithms initially developed for binary classification but posteriorly generalized for multiclass classification or regression analysis. In fact, the regression version of this algorithm has been recently used in Project Management for prediction and project control purposes (Wauters and Vanhoucke, 2014). However, the use of SVM in this paper is focused on classification.

The gist of SVM is founded on a simple binary linear classification algorithm, the maximal margin classifier, which given a set of training examples, each belonging to one of two classes, finds the optimal separating hyperplane that divides the space of features into the two categories maximizing the separation margin between the classes (James et al., 2013). This algorithm classifies new

8       Stochastic Earned Value Analysis using Monte Carlo Simulation and Statistical Learning Techniques.
http://www.sciencedirect.com/science/article/pii/S0263786315001106

observations to the class according to the side of the hyperplane. Although very intuitive, this classifier requires that the categories be linearly separable which is not often the case.

SVM refines this idea extending it to non-linear decision boundaries enlarging the feature space using kernels but also including the concept of soft margin –some training examples are allowed to be in the wrong side of the margin or even the hyperplane– in order to deal with non-separable cases.

One of the problems of SVM comes from the fact that they do not directly give a probability about the prediction, a key feature to control and monitoring support for a project manager. This is usually circumvented fitting a logistic distribution using maximum likelihood to the predicted data of the classifiers (Platt, 2000).

In this paper SVM classifiers with radial kernels have been implemented using R package 'e1071'(Dimitriadou et al., 2008).

### 3.4 As a regression problem

Data obtained from the triad method not only can be used to estimate the probability to incur in overruns. Considering that information as a regression problem, the expected cost and time at termination of the project can also be estimated. In the next sections we give some background about the definition of the regression problem and explain the techniques used in this paper. The validation approach is again nested cross-validation.

#### 3.4.1 Regression

A regression problem entails the prediction of a qualitative or continuous variable, also called *response, output* or *independent variable* as in the classification problem, with a set of qualitative and/or quantitative variables, the *predictors*.

In this section we look to the EVM from another perspective: a regression problem which aims to predict the expected budget and time left at a given time defined by a triad ($x_{AT}$, $T_{xi}$, $C_{xi}$). With this technique, if the project is over-run, we can predict the over-cost and the delay it is experiencing; or the opposite way, how much budget and time the project has left until the finalization. Hence, the expected budget and the expected delay are our *response variables*, and $T_{xi}$ and $C_{xi}$ our *predictors*.

Typically the relationships between the predictors and the response are non-linear in real problems. Although linear models have some benefits such as that allow for relative simple inference, non-linear models are more flexible and may lead to more accurate predictions, at the expense of a less interpretable model. One option to approach this is to include non-linear transformations of the predictors in a linear regression model. In this work we address for an even more flexible alternative, the Generalized Additive Models.

#### 3.4.2 Generalized Additive Models

The Generalized Additive Models (Hastie and Tibshirani, 1990) allow to define non-linear relationships between the predictors and the response, without losing the additive characteristic of the multivariable linear regression models which allow to discern the individual contributions of each predictor to the response. The regression problem is reformulated such that:

$$\hat{Y} = \hat{\beta}_0 + \sum_{j=1}^{p} f_i(X_j) \qquad (2)$$

Where $f_i$ are the unspecified smooth functions. Not all functions need to be non-linear. We can also define nonparametric functions in two predictors, or different functions for each level of a factor (qualitative variable).

In this work we have used two types of flexible representations for $f_i(X)$: natural splines, an expansion of basis functions, and local regression, which belong to the group of scatterplot smoothers.

To fit both regression models, we have used the Backfitting algorithm (Hastie et al., 2009) for fitting additive models estimating all functions simultaneously by iteratively smoothing partial residuals. We have used the R package "gam: Generalized Additive Models" (Wood, 2006).

### 3.4.3 Natural Splines

A smooth function can be represented by using an expansion of basis functions, such as polynomials. A spline is a function piecewise-defined that allows for local polynomial representation, defining several function intervals separated by *knots*. To smooth the places where the polynomial pieces connect, one can impose the function to have continuous first and second derivatives in the knots; this approach is called *cubic spline* and it is the most used type, although higher degree fits can be used if more smoothness is needed in the joints.

A *natural* spline imposes the additional constraint that the function is linear beyond the boundary knots, to avoid the erratic behavior of polynomials near the boundaries (Hastie et al., 2009).

We have used the package "splines" for R. The function ns() generates a natural cubic spline basis matrix and the positions of the knots are adjusted using the observations, given the number of knots.

The model proposed is a GAM using natural splines (*ns*) as functions, as equation (3) summarizes:

$$\hat{Y} = \hat{\beta}_0 + ns(T_{xj}) + ns(C_{xj}) \qquad (3)$$

The number of knots for both predictors has been selected simultaneously so that the model minimizes the Mean Square Error of the predictions, using the nested cross-validation methodology.

### 3.4.4 Local Regression

Local regression is a different approach for fitting a smooth function. It is a nonparametric model which involves fitting a low-degree polynomial model at each point of the training data set, using a subset of the data. The points are weighted so that the closest have the highest weight, using a function to assign these weights, known as *kernel*, and a parameter to define the size of the neighborhood (related to the kernel used), and the model is fitted applying weighted least squares regression. The degree of the polynomials to be used also needs to be defined, typically 1 or 2.

We have used the R package "gam" and the function "lo()" to fit a GAM model with local regression (*loess* from package "stats") as building blocks, as equation (4) summarizes:

$$\hat{Y} = \hat{\beta}_0 + lo(T_{xj}) + lo(C_{xj}) \qquad (4)$$

The degree of polynomials to be fit is set to 1 (default). The size of the neighborhood is controlled by the parameter *span* or α ("for α < 1, the neighborhood includes proportion α of the points, and these have tricubic weighting (proportional to (1 - (dist/maxdist)^3)^3). For α > 1, all points are used, with the 'maximum distance' assumed to be α^(1/p) times the actual maximum distance for p explanatory variables.") (Cleveland et al., 1992). The value of *span* for both predictor models has been chosen from 0.1 to 1 using nested cross-validation and selecting the combination of both span parameters that minimizes the Mean Square Error of the predictions.

## 4. Case study

### 4.1 Description

We illustrate our approach with an example. This selected case has been previously used in project network research (Acebes et al., 2014) and is based on Lambrechts et al. (2008).

The activity-on-node (AON) network is shown in Fig. 3. The network activity durations of this example are modeled as normal distributions, one of the most common in project literature. The chosen network topology highlights the role of parallel paths to better illustrate the usefulness of the approach.

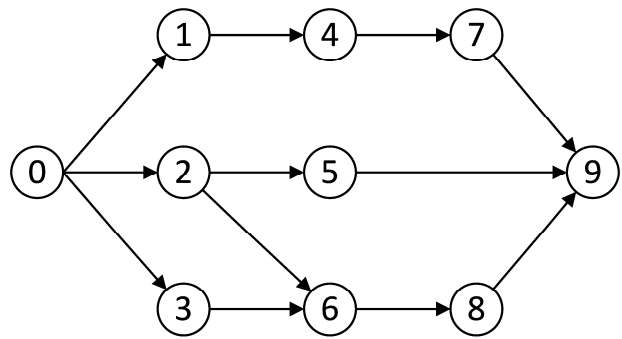

*Fig. 3. AON Network*

The details of the parameters used to model this case are described in Table 1. Please note that for the sake of simplicity, costs are modelled depending deterministically on duration (variable costs determine the cost of the task for time unit). All the methods explained can be used relaxing this assumption and including any stochastic distribution in costs and durations.

*Table 1. Duration and cost of activities of the case study. Duration activities are modelled as normal distributions and costs depend linearly on duration.*

| Id. Activity | Duration | Variance | Variable Cost |
|---|---|---|---|
| A1 | 2 | 0.15 | 755 |
| A2 | 4 | 0.83 | 1750 |
| A3 | 7 | 1.35 | 93 |
| A4 | 3 | 0.56 | 916 |
| A5 | 6 | 1.72 | 34 |
| A6 | 4 | 0.28 | 1250 |
| A7 | 8 | 2.82 | 875 |
| A8 | 2 | 0.14 | 250 |

*In order to calculate the Earned Value of a simulated project, a baseline plan is needed.*

Table 2 shows the PV assumed for the exercise.

*Table 2. Baseline plan for the project*

| Planned Value | 0 | 1 | 2 | 3 | 4 | 5 | 6 | 7 | 8 | 9 | 10 | 11 | 12 | 13 |
|---|---|---|---|---|---|---|---|---|---|---|---|---|---|---|
| A0 | | | | | | | | | | | | | | |
| A1 | | 755 | 755 | | | | | | | | | | | |
| A2 | | 1750 | 1750 | 1750 | 1750 | | | | | | | | | |
| A3 | | 93 | 93 | 93 | 93 | 93 | 93 | 93 | | | | | | |
| A4 | | | | 916 | 916 | 916 | | | | | | | | |
| A5 | | | | | | 34 | 34 | 34 | 34 | 34 | 34 | | | |
| A6 | | | | | | | | | 1250 | 1250 | 1250 | 1250 | | |
| A7 | | | | | | | 875 | 875 | 875 | 875 | 875 | 875 | 875 | 875 |
| A8 | | | | | | | | | | | | | 250 | 250 |
| A9 | | | | | | | | | | | | | | |
| SUM | | 2598 | 2598 | 2759 | 2759 | 1043 | 1002 | 1002 | 2159 | 2159 | 2159 | 2125 | 1125 | 1125 |
| **PV** | | **2598** | **5196** | **7955** | **10714** | **11757** | **12759** | **13761** | **15920** | **18079** | **20238** | **22363** | **23488** | **24613** |

Our analysis is focused when the percentage of work performed measured in terms of earned value of the project is 50% of the budget at completion (planned value at the end of the project), but the same reasoning could be applied to any other percentage.

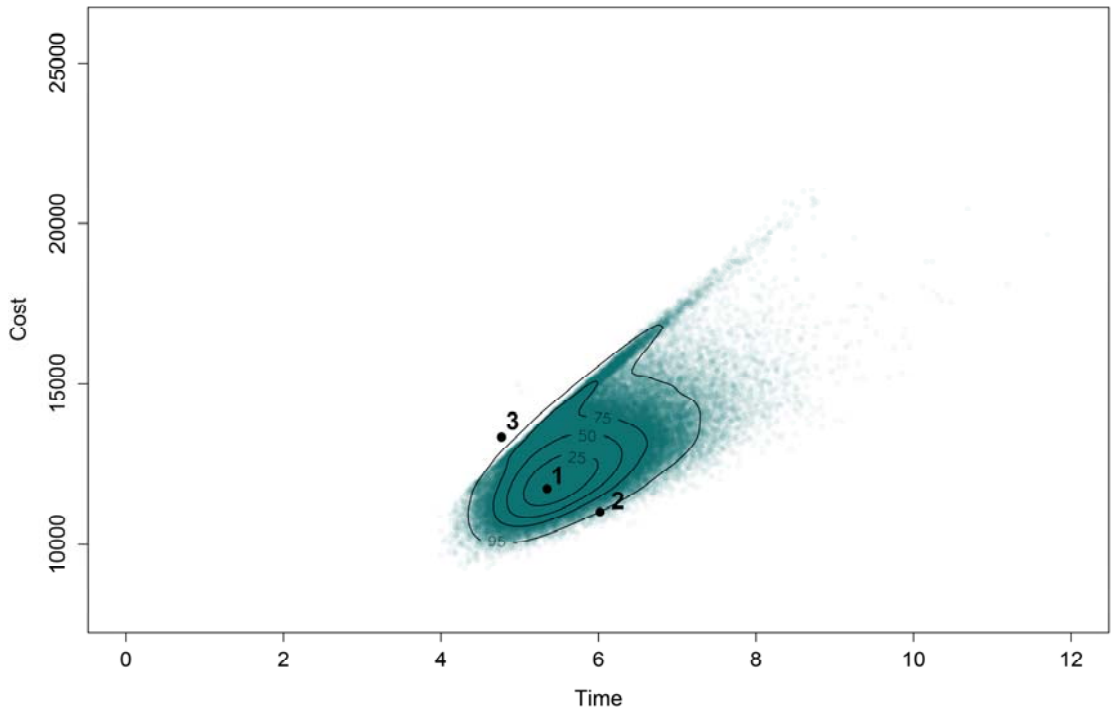

*Fig. 4. 2D probability density estimation obtained for the case study when the EV=50% of BAC. Point 1 represents a point within the expected variability ranges. Point 2 can be considered a warning situation, and the probability that this point is reached just by random is only 5%. Point 3 is a point out of range which previous approaches do not detect as anomalous.*

## 4.2 Results

We begin the analysis with the anomaly detection system obtained through Monte Carlo simulation and posterior 2D density distribution. This analysis allows determining if given an advancement of the project –measured in terms of earned value (corresponding in this case to the 50% of the budget at completion)– , the values of actual time and actual cost of the project are within the expected variability. In Fig. 4 100.000 simulation points of the project for that earned value and an adjusted density distribution that summarizes them are represented. Contour lines represent the probability that the project is without the expected range under the assumption that the project is following the stochastic process considered in the project definition. To illustrate the use, imagine that the project manager decides to control the status of the project at an instant in which the EV is 50% of the BAC, and the actual time and actual cost are given by point 1 in Fig. 4. In this case there is no evidence that suggests any process that is interfering beyond the expected variability of the project. On the contrary, if the point of the project is number 2, the probability that this situation has been obtained as consequence of just randomness is low (5%) and perhaps the project requires to figure out whether there are additional structural causes that are taking place in the project which are deviating it from the planned schedule. Note that this method can detect anomalies consequence of the expected correlation of time and cost (point 3) that may be considered as normal if those variables are decoupled as in previous methods (Acebes et al., 2014). Fig. 5 shows two rectangles computed using the triad methodology, one for the 95% confidence interval (percentiles Pd 2.5 and Pd 97.5 for time, and Pc 2.5 – Pc 97.5 for cost) and another for the 75% confidence interval (Pd 12.5 – Pd 87.5, Pc 12.5 – Pc 87.5). Comparing these predictions with the actual anomaly detection estimation proposed in this work, one can see that, for example, a point in the 95%-square top left area (like point 3) could be erroneously interpreted as a project within the assumed limits of variability using the triad methodology.

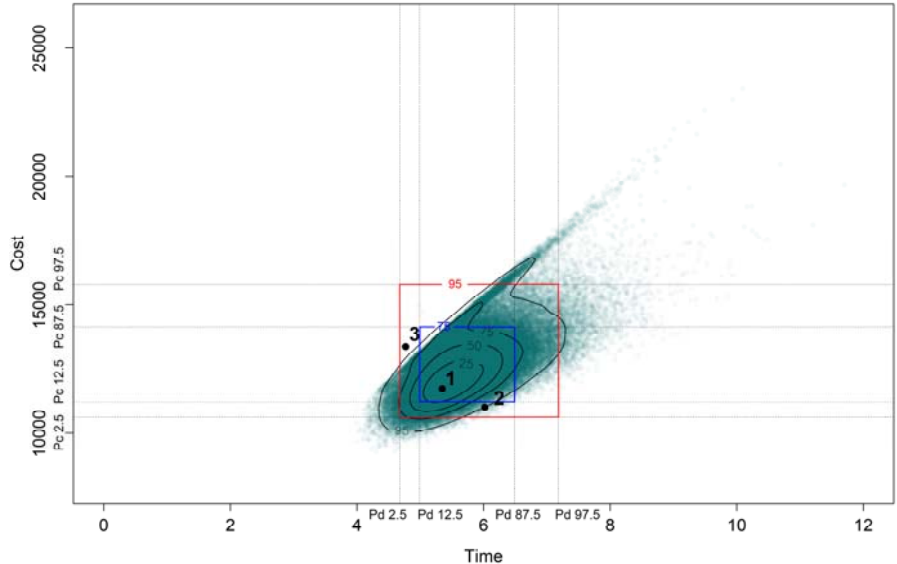

*Fig. 5. Comparison of triad methodology versus anomaly detection. Rectangles represent the 95% and 75% confidence intervals for time and cost using the triad methodology.*

If the conclusion of the previous result is that the model follows the expected variability, we can go beyond in the analysis and determine the probability of finishing the project with delay or overcost depending on the current situation (assuming that the project follows the stochastic pattern defined in the plan). As explained in the methodology section, from complete simulations of the project and pivoting for the same earned value (in our example 50% of BAC), for each pair time and cost, each simulation is classified as a red point (Fig. 6, above on the left) if the simulated project finished with overbudget or green if the project finished with a cost lower than planned. With this dataset several classification models are assessed using nested cross-validation. In the case of our example, SVM with radial kernel obtains better results than the rest of the tested classifiers (although this may depend on the project and the level of development). This model allows to represent the probability of overcost and the decision boundary which splits the points in which is more likely to finish on budget and the points in which is not. Fig. 6 represents with transparency the space where the model gives classification points outside the range the simulation data set, and consequently those predictions should be treated very carefully and in general untrusted.

Analogously, and following the example we have calculated the probability to finish the project in time depending on the current situation. For this process, several classifying algorithms have obtained the same classification rate. We have represented the QDA model for its capacity to capture data correlation. It is interesting to notice that since PV has been calculated using PERT (and consequently is too optimistic) and that there are activities not yet performed with high variance, the perspectives to complete the project without delay are low even in the most favorable situations for this control stage. So much so that in this case there is not any classification boundary.

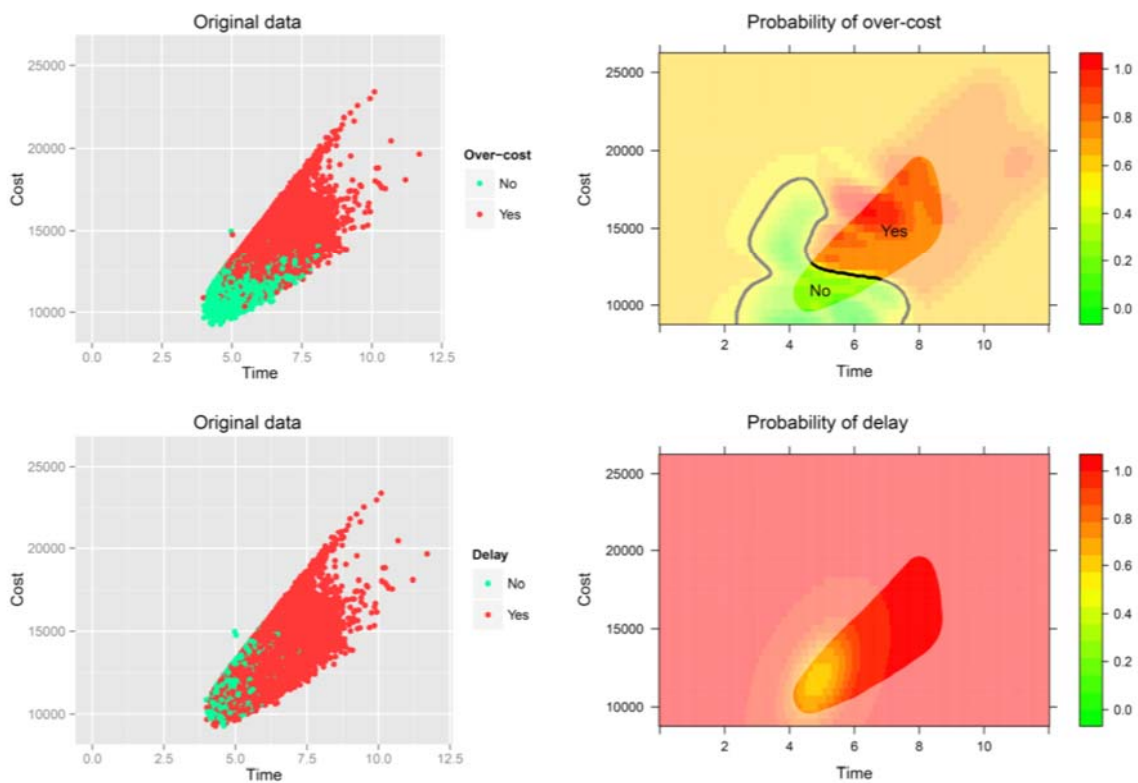

*Fig. 6. Project analysis as a classification problem. Simulation results are labelled as over-run or not depending on the final simulation result (left). These datasets are used to fit classification models (right). These models give an estimation of the probability of over-cost (up) or delay (bottom). Boundary decisions represent points in which the probability of over-run is 0.5.*

Again, our analysis can provide more information to the project manager. It is undoubtedly interesting to estimate the over-run probabilities but it is also relevant to predict their size. From the simulation data and pivoting again for the level of project advancement given by the earned value, for each pair time-cost we can figure out the simulated final duration and total cost of the project (Fig. 7, left figures). This dataset can feed regression models to forecast the expected time and cost of the project if it follows the expected variability. In order to illustrate this process, we have fit two models for each case (time and cost): a generalized additive model with natural splines and a generalized additive model with local regression. An ANOVA test reports that for this case the generalized additive model with natural splines gives better results with a significance of 0.001. The prediction performance of both models can be observed in Fig. 7 where the models have been used to make predictions over a grid and the prediction values have been represented with a heat map (using green color for favorable values and red for problematic ones). The area of data of the grid that is outside the case of study is represented with transparency; again, in this area the predictions of the models may be erroneous.

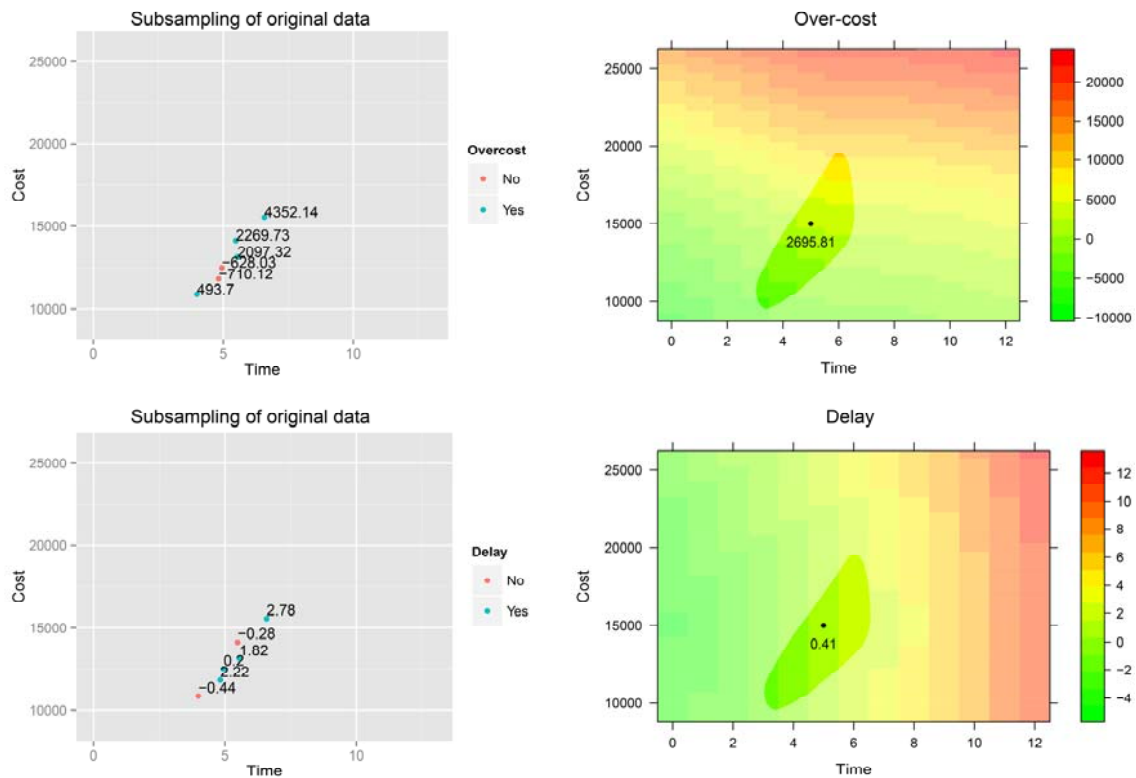

*Fig. 7. Project analysis as a regression problem. Simulation results are labelled with the final simulated cost (up-left) and simulated total duration (bottom-left). These datasets are used to fit regression models (right). These models give an estimation of the probability of the expected over-cost (up-right) or expected delay (bottom-right).*

All this information can be visually integrated in two intuitive graphical control figures (Fig. 8) very similar to the classical Earned Value Management charts. In just two pictures a project manager can obtain not only the popular EVM ratios and indexes but also predictions about the probabilities and expected cost and durations, boundary classifications and the data ranges in which the project is under the expected variability.

At a desired time AT, where Earned Value in the example in Fig. 8 is 50% of BAC (50% EV) and actual cost AC, the project manager can have: the probability that the project is not within the expected variability of the project (represented in Fig. 8 as p(Anomaly)), the probability of over-cost (p(OC), the expected over-cost (negative values represent no over-cost), and the boundary that classifies the point in the "not-expected-over-cost" area (under the line), (these four measures in the top part of Fig. 8); and again the probability of the project in an anomaly situation (p(Anomaly)), the probability of delay (p(D)) and the expected delay (these three measures in the bottom part of Fig. 8).

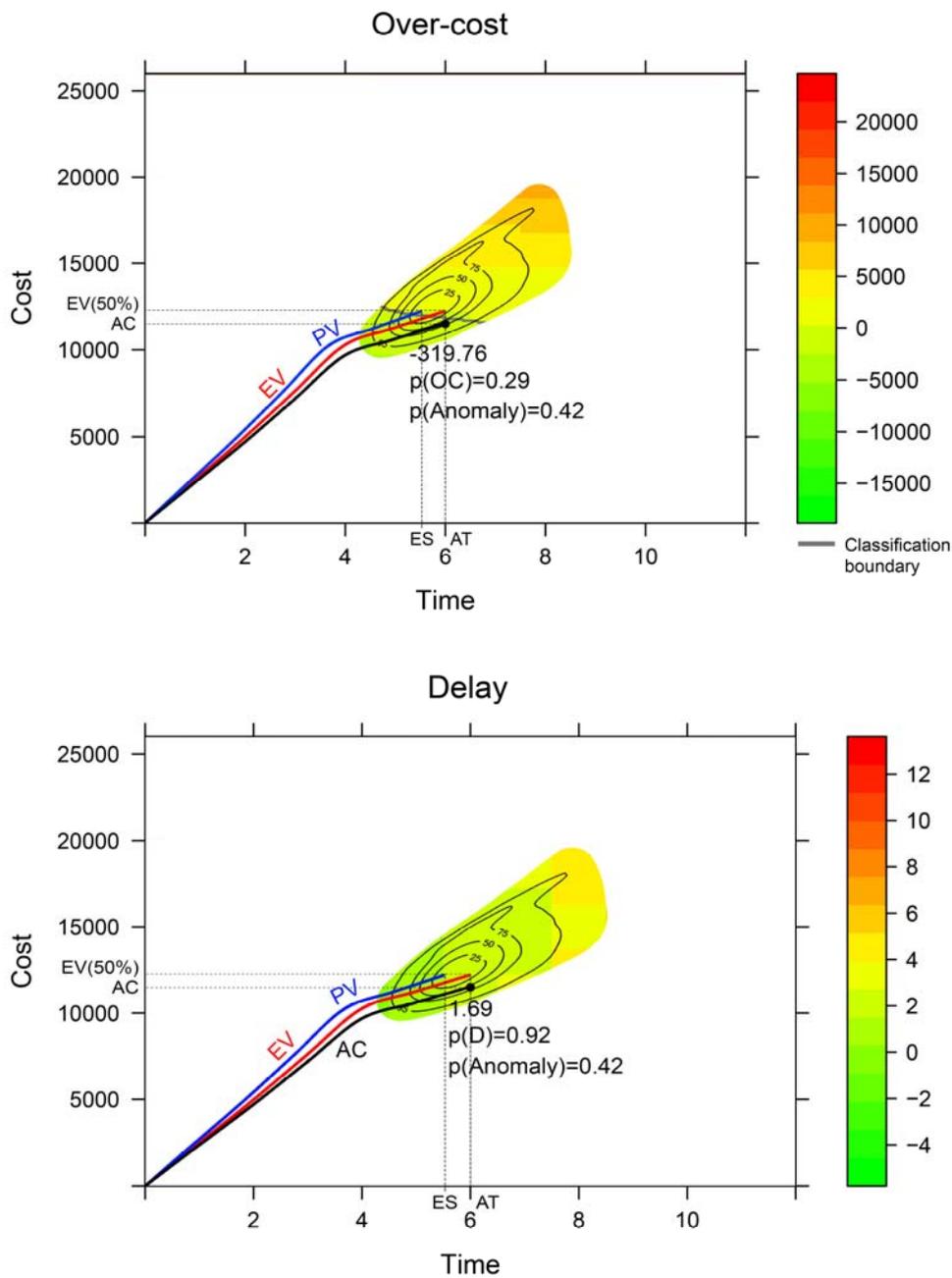

*Fig. 8. Information obtained by the methodology integrated with the classical Earned Value Management graph. The picture on the top represents the expected over-cost, the probability of over-cost, p(OC) and the range of expected variability of the project for the given advancement of the project measured in terms of EV. Analogous measures are represented for the time in the second picture (bottom).*

## 5. Conclusions

In this work we have proposed a refinement for the traditional Earned Value Management method to control projects stochastically modelled. At any stage in the development of the project, the project manager can monitor and control the status of the project. The only data that are needed to feed the algorithm are the stochastic definition of the project, the planned value curve, and the traditional raw calculations of the EVM: EV, AT and AC in that moment. The technique generates multitude of projects compatible with the definition of the project by Monte Carlo simulation. Using

EV as the pivotal measure of the advancement, the project can be analyzed as an anomaly detection, classification and regression problems.

The approach allows detecting anomaly situations in regard to the project definition taking into account the possible correlation between time and cost that previous methodologies ignored. Besides, probabilities of over-runs and the expected time and duration can be also calculated. All this information can be also visually integrated in an intuitive framework compatible with traditional EVM.

No classifier or regression technique is universally better than any other for every possible context, and predicting in advance the relative performance is a challenging task (Bradley, 1997, Hastie et al., 2009). The proposed framework is independent from the algorithms and can be adapted to be used with any other or future detection, classification and regression method. To illustrate the example some of the state of the art techniques have been used, however the approach does not rely on the precise classification or regression algorithm used. On the contrary, we propose the assessment of several techniques and depending on the case to choose the appropriate one for the specific project using cross-validation. Future research may figure out the a priori relationship between the properties of the project (number of tasks, probability distributions used, degree of parallelization of the project, etc) and the prediction results of the different classifiers and forecasting methods, this could reduce the computing time necessary to elaborate the control charts and reports.

## Acknowledgments.

This research has been financed by the project "Computational Models for Strategic Project Portfolio Management", supported by the Regional Government of Castile and Leon (Spain) with grant VA056A12-2 and by the Spanish Ministerio de Ciencia e Innovación Project CSD2010-00034 (SimulPast CONSOLIDER-INGENIO 2010).